\documentclass[a4paper]{jpconf}
\usepackage{graphicx}

\usepackage{ulem}
\usepackage{bm}
\usepackage{color}
\usepackage{amssymb}

\begin{document}
\title{Application of polynomial-expansion Monte Carlo method to a spin-ice Kondo lattice model
}

\author{
Hiroaki Ishizuka$^1$, Masafumi Udagawa$^{1,2}$, and Yukitoshi Motome$^1$
}

\address{$^1$Department of Applied Physics, University of Tokyo, Japan
}
\address{$^2$Max-Planck-Institut f\"{u}r Physik komplexer Systeme, Dresden, Germany
}

\ead{ishizuka@aion.t.u-tokyo.ac.jp
}

\begin{abstract}
We present the results of Monte Carlo simulation for a Kondo lattice model in which itinerant electrons interact 
with Ising spins with spin-ice type easy-axis anisotropy on a pyrochlore lattice.
We demonstrate the efficiency of the truncated polynomial expansion algorithm, 
which enables a large scale simulation, in comparison with a conventional algorithm using the exact diagonalization.
Computing the sublattice magnetization, we show the convergence of the data with increasing the
number of polynomials and truncation distance.
\end{abstract}

\section{Introduction}

Interplay between localized spins and itinerant electrons has been one of the major topics in the
field of strongly correlated electrons.
It triggers various interesting phenomena, for instance, a variety of magnetic orderings induced by
electron-mediated spin interactions, such as Ruderman-Kittel-Kasuya-Yosida (RKKY)
interaction~\cite{RKKYInteraction} and the double-exchange interaction~\cite{DEMechanism}.
The impact of the spin-charge interplay is not limited to the magnetic properties, but also brings 
about peculiar electronic states and transport phenomena, such as non Fermi liquid behavior in the quantum critical region in 
rare-earth systems~\cite{Stewart2001}
and the colossal magneto-resistance in perovskite manganese oxides~\cite{Dagotto2002}.

Recent studies on metallic pyrochlore oxides have opened yet another aspect in the field, namely, the geometrical frustration.
The effect of geometrical frustration on the interplay between itinerant electrons and localized spins 
has attracted much interest, and extensive number of studies on these compounds have been reported.
For example, interesting features were reported in Molybdenum compounds, such as an unconventional anomalous Hall effect~\cite{Taguchi2001}
and emergence of a peculiar diffusive metallic phase~\cite{Iguchi2009}.
A characteristic resistivity minimum was also observed in an Iridium compound~\cite{Nakatsuji2006}.
For these phenomena, the importance of local spin correlations inherent to the strong frustration has been suggested,
but comprehensive understanding is not reached yet.

One of the authors and his collaborator recently reported unbiased Monte Carlo (MC) calculations of a Kondo lattice
model on a pyrochlore lattice~\cite{Motome2010-1,Motome2010-2,Motome2010-3}.
In their studies, however, the accessible system size $N$ was limited to small sizes because the MC simulation
was performed by a conventional algorithm using the exact diagonalization (ED)~\cite{Yunoki1998},
in which the computational amount increases with $O(N^4)$.
Larger size calculations are highly desired to further discuss the peculiar magnetic and transport phenomena in
the frustrated spin-charge coupled systems. 
For this purpose, here we apply another faster algorithm, the polynomial expansion method (PEM)~\cite{Motome1999,Furukawa2004}.
We test the efficiency of the algorithm for a variant of Kondo lattice models on a pyrochlore lattice.

\section{Model and method}

We here consider a Kondo lattice model on a pyrochlore lattice, whose Hamiltonian is given by
\begin{eqnarray}
H = -t \sum_{\langle i,j\rangle, \sigma} \bigl(c_{i,\sigma}^\dagger c_{j,\sigma} + {\rm H.c.}\bigl)
    -J \sum_i {\bf S}_i \cdot \boldsymbol{\sigma}_i.
    \label{eq:H}
\end{eqnarray}
Here, $c_{i,\sigma}$ ($c_{i,\sigma}^\dagger$) denotes an annihilation (creation) operator of electrons
at site $i$ with spin $\sigma (= \uparrow, \downarrow)$; ${\bf S}_i$ and $\boldsymbol{\sigma}_i$ represent
the localized spin and itinerant electron spin, respectively. 
The model is defined on a pyrochlore lattice, three-dimensional frustrated lattice consisting of a corner-sharing network of tetrahedra.
The sum $\langle i,j \rangle$ is taken over the nearest-neighbor (n.n.) sites on the pyrochlore lattice.
We set $t=1$ as the energy unit.

We assume that the localized spins ${\bf S}_i$ are Ising spins with $|{\bf S}_i|=1$, whose 
anisotropy axes depend on the four-sublattice sites on each tetrahedron: the axes are set
along the
$\langle 111 \rangle$ directions, namely, the directions connecting the centers of neighboring tetrahedra. 
The situation is the same as in the spin ice~\cite{Harris1997,Ramirez1999}.
Although there is no bare interaction between the localized spins in the model (\ref{eq:H}), 
the spins communicate with each other through an effective interaction mediated by the kinetic motion of electrons
(RKKY interaction). 
Similar to the spin ice case, when the n.n. interaction is dominantly ferromagnetic (FM), 
a local spin configuration with two spins pointing in and the other two pointing out (two-in two-out) is favored
in each tetrahedron. 
On the other hand, when the interaction is dominantly antiferromagnetic (AFM), all-in or all-out configurations
become energetically stable. 
We note that the sign of the coupling $J$ is irrelevant since the Hamiltonian is unchanged 
for $J \to -J$ and ${\bf S}_i \to -{\bf S}_i$.

We apply MC calculations with PEM to the model (\ref{eq:H}).
In the PEM, the density of states (DOS) for electrons under a spin configuration is obtained using the Chebyshev
polynomial expansion~\cite{Motome1999}, which then is used to calculate the action in the weight for the given spin configuration.
This algorithm reduces the computational cost 
compared to a conventional MC method based on ED~\cite{Yunoki1998}. 
We also implement the truncation method which further reduces the computational cost~\cite{Furukawa2004}. 
We carry out the truncation by a real space distance, not by a magnitude of the matrix element in the original scheme;
namely, we introduce a truncation distance defined by the Manhattan distance from a flipped spin 
in calculating the Chebyshev moments. 
The total computational amount for each MC step is largely reduced from $O(N^4)$ for the conventional algorithm to
$O(N)$~\cite{Furukawa2004}.
This enables us to access to larger system sizes.

The calculations are done with varying the number of polynomials $15 \le m \le 50$ and truncation distance $3 \le d \le 7$.
In the following, we take $J=2$ and restrict ourselves to two typical electron densities; a low electron density
$n_e = \sum_{i,\sigma} \langle c_{i,\sigma}^\dagger c_{i,\sigma} \rangle / N \sim 0.03$
(the chemical potential is fixed at $\mu = -5.9$), for which the n.n.
RKKY interaction is FM, and an intermediate electron density $n_e \sim 0.35$ ($\mu = -1.3$),
for which the n.n. RKKY interaction is AFM.
%All the following calculations were done at $J=2$ for $N=4\times 4^3$ with periodic boundary conditions,
All the following calculations were done %at $J=2$
for $N=4\times 4^3$ with periodic boundary conditions, typically with 2700 MC steps after 500 steps of thermalization.
The results are divided into three bins of 900 steps to estimate the statistical error with 100 steps intervals in between each bin.
MC simulation %with $m=40$ and $d=6$, for instance,
costs about 15 hours with 8 cpu parallelization of Intel Xeon E5502 processors~\cite{note}. 

\section{Results}

Here, we show the MC results for the square of the sublattice magnetization, $m_s^2$, with different $m$ and $d$.
We calculate $m_s^2$ by the diagonal component of the spin structure factor 
$
S^{\alpha\beta}({\bf q}) =
\sum_{n,l} \langle {\bf S}_{n}^\alpha {\bf S}_{l}^\beta \rangle \exp ({\rm i}{\bf q}\cdot {\bf r}_{nl}) / N
$, as $m_s^2 = 4 S^{\alpha\alpha}({\bf q}={\bf 0}) / N$.
Here, $n$ and $l$ are the indices of tetrahedra and $\alpha,\beta=1,2,3,4$ are the indices of sites in a tetrahedron.

\begin{figure}
  \begin{center}
  \begin{minipage}{4.1in}
  \vspace{0.3in}
  \includegraphics[width=4.2in]{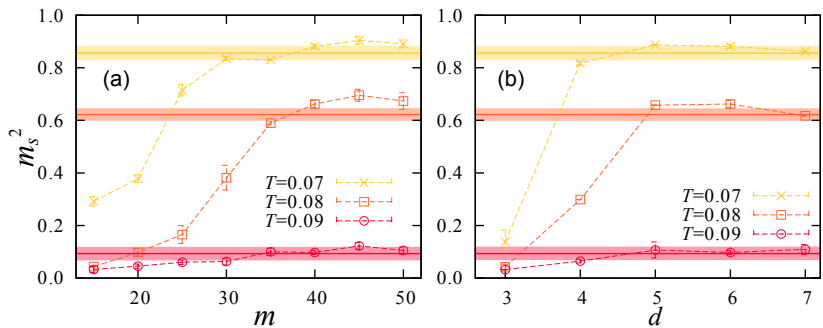}
  \end{minipage}
  \hspace{0.1in}
  \begin{minipage}{2.0in}
  \caption{
  MC results by the PEM for $m_s^2$ at an intermediate density $n_e \sim 0.35$.
  We take $d=6$ in (a) and $m=40$ in (b). 
  The results and errors of the ED method are shown by horizontal solid lines and shades.
  }
  \label{fig:all}
  \end{minipage}
  \end{center}
\end{figure}

Figure~\ref{fig:all} shows the PEM results of $m_s^2$ at an intermediate density $n_e \sim 0.35$ for three typical temperatures($T$).
Figure~\ref{fig:all}(a) shows $m$ dependence at $d=6$ and 
figure~\ref{fig:all}(b) is for $d$ dependence at $m=40$.
At this electron density, the localized spins align in the all-in/all-out configuration
(alternative arrangement of all-in and all-out tetrahedra) at low $T$ since the n.n. RKKY interaction is dominantly AFM. 
The horizontal solid lines denote the results obtained by the ED MC method.
The ED result at $T=0.07$ shows $m_s^2 > 0.8$, which implies the system to be in the AFM ordered phase.
At $T=0.09$, $m_s^2 \sim 0.1$, which indicates a disordered paramagnetic state.
The intermediate $T=0.08$ is presumed to be around the critical point.
In figure~\ref{fig:all}(a), the PEM results at $T=0.07$ and $T=0.09$ converge to the ED results when $m \gtrsim 30$.
The data at $T=0.08$, however, show slower convergence: It appears to require $m \gtrsim 35\sim 40$ for reliable calculation.
The convergence as to $d$ shows a similar tendency, as shown in figure~\ref{fig:all}(b); 
$d \gtrsim 4$ is sufficient at $T=0.07$ and $T=0.09$, while a larger $d \gtrsim 5$ appears to be necessary at $T=0.08$.

In lower density regions, the convergence becomes poorer because of several reasons. 
One is simply because the relevant $T$ range becomes lower; 
the RKKY energy scale gets smaller for lower density. 
Another reason is that the Fermi energy comes close to the band edge (bottom); 
the PEM is an expansion technique of DOS, and hence, larger $m$ is necessary 
for good convergence when DOS changes rapidly as in the band edge. 
Furthermore, when the electron density is small, a fluctuation of the density in MC measurements will harm the precision;
%in calculations of physical quantities 
in our calculations, $n_e$ has typically an error of $\Delta n_e \simeq 0.01$, which might affect the results when $n_e \sim \Delta n_e$.

\begin{figure}
  \begin{center}
  \begin{minipage}{4.1in}
  \vspace{0.3in}
  \includegraphics[width=4.2in]{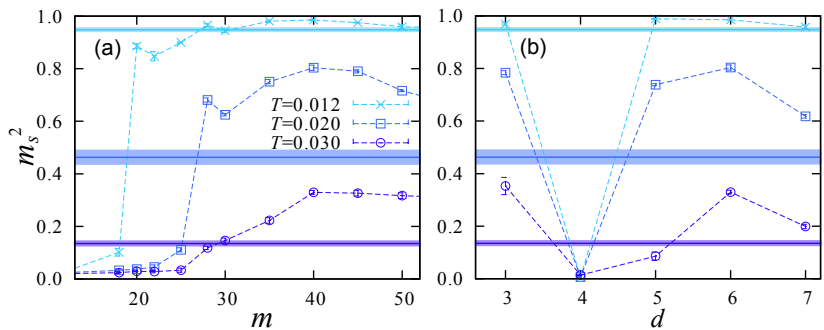}
  \end{minipage}
  \hspace{0.1in}
  \begin{minipage}{2.0in}
  \caption{
  MC results by the PEM for $m_s^2$ at a low density $n_e \sim 0.03$.
  We take $d=6$ in (a) and $m=40$ in (b).
  The results and errors of the ED method are shown by horizontal solid lines and shades.
  }
  \label{fig:ice}
  \end{minipage}
  \end{center}
\end{figure}

Such situation is illustrated in figure~\ref{fig:ice} for an extremely low density $n_e \sim 0.03$.
In this region, the lowest $T$ state is characterized by a FM order of two-in
two-out tetrahedra with aligning the net moment of each tetrahedra along a $\langle 100 \rangle$ direction.
Note that the $T$ range in figure~\ref{fig:ice} is much smaller than in figure~\ref{fig:all}. 
The data show much poor convergence to the ED results: 
Nevertheless, $m \gtrsim 40$ and $d \gtrsim 7$ will give converged results even in this extreme case.

Our result indicates that reasonable convergence to the ED result is achieved by taking 
$m \sim 35-40$ and $d \sim 5-7$ in the PEM in a wide range of the electron density and temperature. 
The convergence is considerably slower than in the previous study for a double-exchange model
on the three-dimensional cubic lattice~\cite{Motome2003}, which showed sufficient convergence within $m \le 8$.
This is partly due to the difference of the magnitude of $J$.
In the previous study, $J$ was taken to be infinity, which greatly simplifies the band structure, 
while we take a much smaller value $J=2$ in the present study. 
The resultant complicated band structure requires a larger number of $m$ to reproduce its fine details.
Another possibility is the effect of geometrical frustration in the present model.
In general, the frustration leads to degeneracy among different spin configurations in a small energy window. 
For such situation, a small systematic error in the calculation of the MC weight could lead to considerable errors in observables.
Furthermore, the frustration suppresses all the energy scales, which might also require much larger efforts to obtain converged data.

\section{Summary}
To summarize, we examined the efficiency of the Monte Carlo simulation using the polynomial expansion method for
a frustrated Kondo lattice model with spin-ice type Ising spins.
By comparison with the results by the exact diagonalization method, our result indicated that the polynomial-expansion Monte Carlo
calculation with $m \sim 35-40$ and $d \sim 5-6$ gives consistent results in a wide range of electron density and temperature.
Larger $m$ and $d$ are necessary for convergence near the critical temperature and for very low electron density.
Our results demonstrate that the polynomial expansion method is a practical method for investigating
the physics of spin-charge coupled systems even when $J$ is much smaller than the bandwidth.
Monte Carlo study of the phase diagram by systematic analysis up to larger system sizes is in progress. 

The authors thank T. Misawa for fruitful discussions. H.I. is supported by Grant-in-Aid for JSPS Fellows.
This work is supported by KAKENHI (Grants No. 19052008, 21340090, 21740242, and 22540372),
the Global COE Program ``the Physical Sciences Frontier", and HPCI Strategic Program, from MEXT, Japan.


\begin{thebibliography}{99}
\bibitem{note} The computational cost is even larger compared to the ED method for this system size, but the relative cost gets much smaller very quickly with increasing $N$.
\end{thebibliography}
\end{document}